\documentclass[aps,prb,twocolumn,showpacs,superscriptaddress]{revtex4}

\usepackage{graphicx} 

\begin{document}

\title
{Confined states in multiple quantum well structures of
Si$_{n}$Ge$_{n}$ nanowire superlattices}
\author{N. Akman}
\affiliation{Department of Physics, Mersin University, Mersin
33343, Turkey}
\author{E. Durgun}
\affiliation{Department of Physics, Bilkent University, Ankara
06800, Turkey} \affiliation{UNAM-Institute for Materials Science
and Nanotechnology, Bilkent University, Ankara 06800, Turkey}
\author{S. Cahangirov}
\affiliation{UNAM-Institute for Materials Science and
Nanotechnology, Bilkent University, Ankara 06800, Turkey}
\author{S. Ciraci} \email{ciraci@fen.bilkent.edu.tr}
\affiliation{Department of Physics, Bilkent University, Ankara
06800, Turkey} \affiliation{UNAM-Institute for Materials Science
and Nanotechnology, Bilkent University, Ankara 06800, Turkey}
\date{\today}

\begin{abstract}
Mechanical properties, atomic and energy band structures of bare
and hydrogen passivated Si$_{n}$Ge$_{n}$ nanowire superlattices
have been investigated by using first-principles pseudopotential
plane wave method. Undoped, tetrahedral Si and Ge nanowire
segments join pseudomorphically and can form superlattice with
atomically sharp interface. We found that Si$_{n}$ nanowires are
stiffer than Ge$_{n}$ nanowires. Hydrogen passivation makes these
nanowires and Si$_{n}$Ge$_{n}$ nanowire superlattice even more
stiff. Upon heterostructure formation, superlattice electronic
states form subbands in momentum space. Band lineups of Si and Ge
zones result in multiple quantum wells, where specific states at
the band edges and in band continua are confined. The electronic
structure of the nanowire superlattice depends on the length and
cross section geometry of constituent Si and Ge segments. Since
bare Si and Ge nanowires are metallic and the band gaps of
hydrogenated ones varies with the diameter, Si$_{n}$Ge$_{n}$
superlattices offer numerous alternatives for multiple quantum
well devices with their leads made from the constituent metallic
nanowires.
\end{abstract}


\maketitle

\section{introduction}
Planar superlattices have been fabricated either through periodic
junction of alternating semiconductor layers with different band
gaps or through repeating compositional modulation. Electrons in
parallel layers show two-dimensional (2D) free electron-like
behavior and have quantization different from those 3D bulk
semiconductors. Minibands in the momentum space in the direction
perpendicular to the layers and periodically varying band gap in
the direct space have attributed unusual electronic functions for
novel devices. These devices are field effect transistors,
photodetectors, light emitting diodes (LEDs) and quantum cascade
lasers etc.

Recently, new growth techniques have enabled also the synthesis of
one-dimensional (1D) nanowire superlattices (NWSLs). NWSLs from
group III-V and group IV elements have been synthesized
successfully. InAs/InP superlattices \cite{Bjork} with atomically
perfect interfaces and with periods of several nanometers could be
realized using techniques, such as molecular beam epitaxy and
nanocluster catalyst. Furthermore, compositionally modulated
superlattices of GaAs/GaP have been synthesized by laser-assisted
catalytic growth technique\cite{Gudiksen} again with atomically
perfect interfaces and with the component layers ranging from 2 to
21. It is proposed that these NWSL's can offer potential
applications in nanoelectronics such as optical nanobar codes, 1D
waveguides and polarized nanoscale LEDs. Longitudinal Si/Si-Ge
NWSL with nanowire diameter ranging from 50 to 300 nm have been
also synthesized using laser ablation growth technique.\cite{Wu}
Structural parameters such as nanowire diameter, Ge concentration
and the modulation period in the Si/Si-Ge superlattices can be
controlled easily by adjusting the reaction conditions.
Technological applications such as LEDs and thermoelectric devices
have been suggested. In addition to the longitudinal (axial)
nanowire superlattices, coaxial core-shell and core-multishell
nanowire heterostructures have attracted interest recently.
Crystalline Si/Ge and Ge/Si core-shell structures have been
experimentally synthesized by Lauhon {\it et al}.\cite{Lauhon}
Most of works involved in 1D superlattices especially concern the
experimental synthesis and characterization of coaxial nanowire
heterostructures.

Theoretically, only a few works investigated core-shell and
longitudinal NWSL's.\cite{Musin, Kagimura, Voon, Zypman} Kagimura
{\it et al}\cite{Kagimura} reported an {\it ab initio} study of
the electronic properties of Si and Ge nanowires, Si/Ge
heterostructures with one surface dangling bond state per unit
cell. They concluded that surface dangling bond level observed in
the band gap of nanowires and nanowire heterostructures can be
used as reference level to estimate band lineups in these systems.
Using one-band effective mass theory, a criterion has been
developed for the occurrence of longitudinal barrier
height.\cite{Voon} It has been argued that radial confinement
reduces the actual barrier height in modulated nanowire
superlattices. Zypman\cite{Zypman} used the Hubbard model to get
the energy spectrum of one-dimensional systems and applied their
results to various model systems like nanowire tunnelling diodes
and Si/Ge superlattice nanowires to interpret the scanning
tunnelling spectroscopy measurements. Earlier formation of
multiple quantum well structure and resulting confined states on
hydrogenated or radially deformed carbon nanotubes have been also
reported.\cite{Salim1}

In this paper, we have investigated mechanical properties, atomic
and electronic structures of bare and hydrogenated
Si$_{n}$Ge$_{n}$ nanowire superlattices using first-principles
plane wave method. NWSL's are constructed from alternating Si and
Ge nanowire segments (zones), both have same orientation and
similar atomic structure. These segments are joined
pseudomorphically and formed a sharp interface. We found that even
a small diameter hydrogenated Si$_{n}$Ge$_{n}$ NWSL's form
multiple quantum well structures where conduction and valence band
electrons are confined. Our study indicates that the band lineup
and resulting electronic structure depend on the length and cross
section geometry of the constituent Si$_{n}$ and Ge$_{n}$
nanowires.

\section{Method}
We have performed first-principles plane wave
calculations\cite{payne, vasp} within DFT\cite{kohn} using
ultra-soft pseudopotentials.\cite{vasp, vander} The exchange
correlation potential has been approximated by generalized
gradient approximation GGA using PW91 functional.\cite{gga} For
partial occupancies we use the Methfessel-Paxton smearing
method.\cite{methfessel} The adopted smearing width is 0.1 eV for
the atomic relaxation and 0.02 for the accurate band structure
analysis and density of state calculations. All structures have
been treated within a supercell geometry using the periodic
boundary conditions. The lattice parameters of the tetragonal
supercell are $a_{sc}$, $b_{sc}$ and $c_{sc}$. We took
$a_{sc}=b_{sc}$=27 \AA~ for NWSL having the largest diameter
($\sim$ 1.8 nm), but $a_{sc}=b_{sc}$=22 \AA~ for one having the
smallest diameter ($\sim$ 1.2 nm) considered in this paper. These
values allowed minimum distance ranging from $\sim$ 11 \AA~ to 14
\AA~ between two atoms in different adjacent cells, so that their
coupling is hindered significantly. We took $c_{sc}$ equal to the
lattice constant $c$ of the nanowires and NWSLs under
consideration. In the self-consistent potential and total energy
calculations the Brillouin zone (BZ) is sampled in the
\textbf{k}-space within Monkhorst-Pack scheme\cite{monk} by
(1x1x9) mesh points for single unit cell and for example (1x1x5)
mesh points for double cells. A plane-wave basis set with kinetic
energy up to 250 eV has been used. All atomic positions and
lattice constant $c_{sc}=c$ are optimized by using the conjugate
gradient method where total energy and atomic forces are
minimized. The criterion of convergence for energy is chosen as
10$^{-5}$ eV between two ionic steps, and the maximum force
allowed on each atom is 0.05 eV/$\AA$.

\begin{center}
\begin{figure}
\includegraphics[scale=0.4]{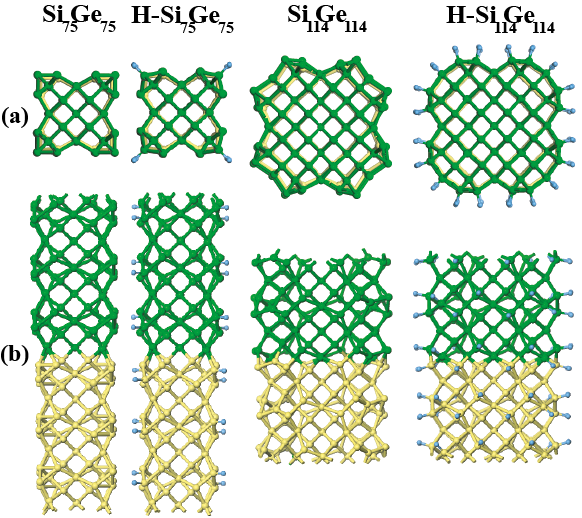}
\caption{(Color online) Optimized atomic structure of bare and
hydrogenated Si$_{n}$Ge$_{n}$ nanowire superlattices for {\it
n}=75 and 114. a) Top view; b) Side view. Small-gray, large-light
and large-dark balls correspond to the hydrogen, silicon and
germanium atoms respectively. }\label{fig:structure}
\end{figure}
\end{center}

\section{Bare and Hydrogenated Nanowires and Nanowire Superlattices}
In this study we considered bare and hydrogen passivated
longitudinal Si$_{n}$Ge$_{n}$ nanowire superlattices and also bare
and hydrogen passivated Si and Ge nanowires as constituent
structures. Bare Si and Ge nanowires are oriented along [001]
direction of the parent diamond crystal and have normally {\it N}
atoms in their primitive unit cell with lattice constant $c$ along
the nanowire (or z-) axis. We took {\it N}=25 and {\it N}=57 as
two special prototypes. We designate them as SiNW({\it n})
[GeNW({\it n})] or shortly Si$_{n}$ (Ge$_{n}$) with {\it n=sN},
{\it s} being an integer number. Si$_{n}$ and Si$_{N}$ (Ge$_{n}$
and Ge$_{N}$) indicates the same nanowire, except that the unit
cell of the former one includes {\it s} primitive cell in direct
space with {\it 1/s} times reduced BZ in the momentum space. In
our simulations bare Si$_{n}$ (Ge$_{n}$) nanowires are first cut
from the bulk crystal with ideal structural parameters.
Subsequently ideal bare nanowires are relaxed to optimize their
structure and lattice constant. Si (Ge) atoms near the core of
relaxed nanowire has tetrahedral coordination. To obtain
H-passivated Si$_{n}$ or Ge$_{n}$ nanowires (designated as
H-SiNW({\it n}) or H-GeNW({\it n}), shortly as H-Si$_{n}$ or
H-Ge$_{n}$) the dangling bonds at the surface are saturated by H
atoms and whole structure is re-optimized. Our study indicates
that the atomic and electronic structure of H-Si$_{n}$ and
H-Ge$_{n}$ may depend on whether hydrogen passivation and
subsequent optimization are achieved on ideal or optimized bare
Si$_{n}$ and Ge$_{n}$ nanowires. The present sequence of structure
optimization mimics the actual growth of hydrogen passivated
nanowires.

A Si$_{n}$Ge$_{n}$ has {\it n=sN} Si atoms at one side and {\it
n=sN} Ge atoms at the other side of NWSL unit cell. These atoms
have tetrahedral coordination as if they are part of a SiGe
heterostructure and hence at the interface Si atoms are bonded to
Ge atoms pseudomorphically and make atomically flat interface. We
note that pseudomorphic growth can sustain for small diameters;
but misfit dislocations may be generated at the interface of large
diameter (or large {\it N}) Si$_{n}$Ge$_{n}$ superlattice. Atomic
positions and lattice constant are relaxed to obtain optimized
structure. H-Si$_{n}$Ge$_{n}$ follow the same sequence of
construction as H-Si$_{n}$ or H-Ge$_{n}$. Optimized lattice
constants of bare Si$_{n}$Ge$_{n}$ nanowire superlattice for {\it
n}=25, 50 and 75 are found to have $c$=10.9 \AA, 21.8 \AA~ and
32.7 \AA, respectively. Upon hydrogenation these lattice constants
change to $c$=11.2 \AA, 22.3 \AA~ and 33.5 \AA, respectively.
Lattice constants of bare and hydrogenated Si$_{n}$Ge$_{n}$, {\it
n}=57 and 114 are almost identical and are $c$=11.1 \AA~ and 22.2
\AA, respectively. Fig.\ref{fig:structure} shows the atomic
structure of bare and hydrogen passivated Si$_{n}$Ge$_{n}$ for
{\it n}=75 and 114. These NWSLs are reminiscent of
Si$_{n}$Ge$_{n}$ (001) planar superlattice which were fabricated
by molecular beam epitaxy by growing first {\it n} Si (001) plane
and then {\it n} Ge (001) plane, and eventually by repeating this
Si$_{n}$Ge$_{n}$ (001) unit periodically. While the
Si$_{n}$Ge$_{n}$ (001) superlattice has 2D periodicity in (001)
layers, NWSLs under study here have finite cross section and hence
2D periodicity is absent. Electrons are bound to NWSL in radial
(lateral) direction, but propagate as 1D Bloch states along the
superlattice axis (in longitudinal direction).

\begin{center}
\begin{figure}
\includegraphics[scale=0.43]{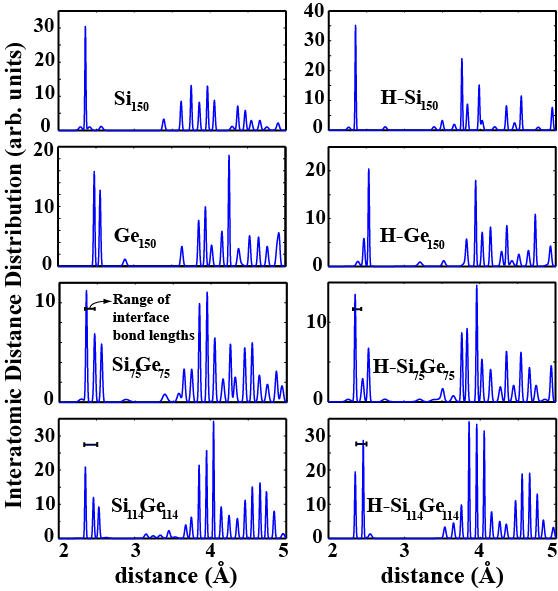}
\caption{(Color online) Interatomic distance distribution of
optimized bare and hydrogenated Si$_{2n}$, Ge$_{2n}$ and
Si$_{n}$Ge$_{n}$ for {\it n}=75 up to fourth nearest neighbor.
The similar distribution for Si$_{114}$Ge$_{114}$ and
H-Si$_{114}$Ge$_{114}$ are also shown. The Si-H (Ge-H) bond
lengths being in the range of ($\sim$ 1.5 \AA) are not
shown.}\label{fig:bond_pops}
\end{figure}
\end{center}

Interatomic distance distribution of Si$_{75}$Ge$_{75}$ and
H-Si$_{75}$Ge$_{75}$ NWSLs are compared with parent Si and Ge
nanowires in Fig.\ref{fig:bond_pops}. In the same figure we also
show the interatomic distance distribution of bare and
hydrogenated Si$_{114}$Ge$_{114}$ NWSL. At the surface, optimized
atomic structures of Si$_{n}$ and Ge$_{n}$ deviate considerably
from the ideal structure of Si$_{n}$ and Ge$_{n}$. For example,
one can deduce quadrangles of atoms at the surface. Normally,
NWSLs consist of hexagonal and pentagonal rings, where one can
distinguish bond lengths in different categories. The interatomic
distance distribution of Si$_{n}$Ge$_{n}$ is reminiscent of the
sum of those of Si$_{2n}$ and Ge$_{2n}$ except some changes
originated from the interface between Si and Ge segments of
supercell. While bulk optimized Si-Si and Ge-Ge bond lengths are
{\it d}=2.36 \AA, and 2.50 \AA, respectively, the Si-Ge bond at
the interface ranges between 2.35 \AA~ and 2.52 \AA~ for bare
Si$_{228}$Ge$_{228}$ (between 2.37 \AA~ and 2.49 \AA~ for
H-Si$_{228}$Ge$_{228}$). Nevertheless, the distribution exhibit
several peaks corresponding to the deviations from the bulk
geometry at the surface. As the cross section or {\it N} increases
the effect of the surface decreases and the distribution of
interatomic distances becomes more bulk-like.

\section{mechanical properties}
The stability and elasto-mechanical properties of Si$_{n}$Ge$_{n}$
and H-Si$_{n}$Ge$_{n}$ NWSLs are crucial for their possible use in
nanoelectronics. In the present study the maximum diameter of
nanowire we treated is $\sim$ 1.8 nm. The diameter of hydrogenated
Si$_{25}$Ge$_{25}$ NWSL is even smaller ($\sim$ 1.4 nm). For such
small diameter nanowires or NWSL's, there are ambiguities in
determining the area of cross section. Moreover, the surface to
volume ratio is rather high and hence makes the cross section
nonuniform. In view of these, the calculation of Young's modulus
may not be appropriate. Here we rather considered the force
(spring) constants of nanowires and NWSLs under a strain in the
harmonic region. To this end we calculated the second derivative
of the total energy (per unit cell) with respect to the lattice
constant $c$ ({\it i.e.} $\kappa$=$d^{2} E_{T}$/$dc^{2}$) or to
the strain, $\epsilon$=$\Delta c/c$ ({\it i.e. }
$\kappa^{\prime}$=$d^{2} E_{T}$/$d\epsilon^{2}$). The values
calculated for nanowires and NWSLs treated in our paper are given
in Table \ref{table:elastic}.

\begin{table}
\caption{Equilibrium values of lattice parameter {\it c} are given
in units of \AA. Force constant $\kappa$ (as defined in the text), in units of eV/\AA, is
calculated by using both VASP result and Hook's law. Percentage
difference in between force constant values calculated from VASP
result and Hook's law is given within parenthesis in order to
check whether classical Hook's law is still valid in nanoscale.
Also force constant $\kappa^{\prime}$ (as defined in the text)
is presented in units of eV. } \centering
\label{table:elastic}
\begin{tabular}{c|ccccc}
\hline
\multicolumn{1}{c|}{structure}&\multicolumn{1}{c}{$c_{o}$}&\multicolumn{1}{c}{$\kappa$}&
\multicolumn{1}{c}{Hook's Law}&\multicolumn{1}{c}{}&\multicolumn{1}{c}{$\kappa^{\prime}$}\\
\hline\hline
Si$_{25}$&5.32&5.68&&&161\\
Ge$_{25}$&5.57&3.28&&&102\\
Si$_{25}$Ge$_{25}$&10.90&2.18&2.08&(5)&259\\
Si$_{50}$Ge$_{50}$&21.75&0.92&1.04&(12)&437\\
Si$_{75}$Ge$_{75}$&32.70&0.62&0.69&(10)&663\\
Si$_{57}$&5.43&11.22&&&327\\
Ge$_{57}$&5.65&7.49&&&239\\
Si$_{57}$Ge$_{57}$&11.07&4.24&4.49&(6)&522\\
Si$_{114}$Ge$_{114}$&22.15&2.10&2.25&(7)&1035\\
H-Si$_{25}$&5.45&8.56&&&254\\
H-Ge$_{25}$&5.73&5.98&&&196\\
H-Si$_{25}$Ge$_{25}$&11.17&3.48&3.52&(1)&436\\
H-Si$_{50}$Ge$_{50}$&22.30&1.70&1.76&(3)&845\\
H-Si$_{75}$Ge$_{75}$&33.50&1.14&1.17&(3)&1279\\
H-Si$_{57}$&5.39&13.57&&&394\\
H-Ge$_{57}$&5.68&11.09&&&358\\
H-Si$_{57}$Ge$_{57}$&11.05&6.13&6.10&(1)&755\\
H-Si$_{114}$Ge$_{114}$&22.08&3.33&3.05&(8)&1626\\
\hline
\end{tabular}
\end{table}

Like bulk crystals, Si$_{n}$ nanowires are stiffer than Ge$_{n}$
nanowires. This implies that the lattice mismatch between Si and
Ge nanowires in NWSL is accommodated mainly by the Ge zone. For
both nanowires and NWSL, $\kappa$ increases with increasing cross
section. For example $\kappa$ of Si$_{25}$ is almost the half of
$\kappa$ of Si$_{57}$. Note that $\kappa$(Si$_{50}$) $\simeq$
$\kappa$(Si$_{25}$)/2. As for, $\kappa$ of Si$_{25}$Ge$_{25}$ NWSL
calculated from first principles is 2.18 eV/\AA. This value can be
estimated in terms of two springs connected in series, namely
$\kappa^{-1}$(Si$_{25}$Ge$_{25}$) $\simeq$
$\kappa^{-1}$(Si$_{25}$)+$\kappa^{-1}$(Ge$_{25}$) to be
$\kappa$(Si$_{25}$Ge$_{25}$) $\simeq$ 2.08 eV/\AA. We, therefore,
conclude that as long as the geometry and size of the cross
section remained to be similar, classical Hook's law continues to
be approximately valid even for nanostructures. Upon hydrogenation
both nanowires as well as NWSLs studied here become stiffer. The
spring constant of Si$_{57}$Ge$_{57}$ is twice that of
Si$_{114}$Ge$_{114}$, because the latter NWSL has twice the length
of the former. We also calculated \cite{strain} the ratio of the
strain of the Ge-zone to that of Si-zone of Si$_{75}$Ge$_{75}$
under tensile stress, {\it i.e.} $\epsilon$ (Ge)$/\epsilon$ (Si)
to be $\sim$ 2.5. This ratio is reduced to $\sim$ 1.25 for
Si$_{114}$Ge$_{114}$. In compliance with the $\kappa$ values in
Table \ref{table:elastic}, this result indicates that in a
Si$_{n}$Ge$_{n}$ NWSL Ge zone elongates more than Si-zone. Using
empirical potential, Menon {\it et al}\cite{Menon} was able to
calculate the Young's modulus and bending stiffness of tetrahedral
and cage-like Si nanowire of $\sim$ 4 nm diameter, and found
values comparable with bulk values.

\begin{center}
\begin{figure}
\includegraphics[scale=0.53]{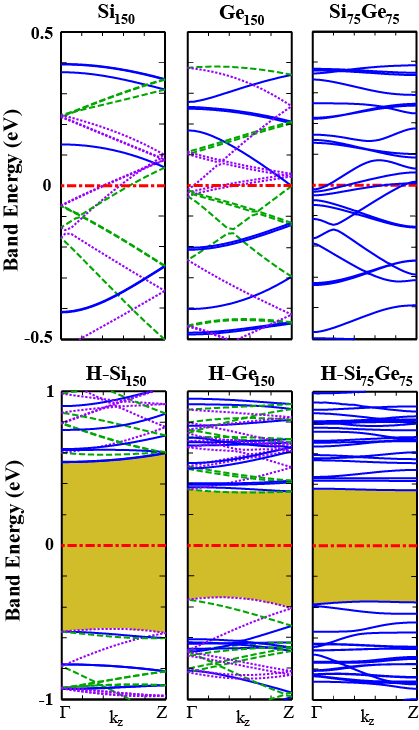}
\caption{(Color online) Energy band structures of optimized bare
and hydrogenated Si$_{2n}$, Ge$_{2n}$ nanowires and
Si$_{n}$Ge$_{n}$ nanowire superlattices for {\it n}=75. Zero of
energy is taken at the Fermi level. Band gaps are shown by shaded
zones. Dashed and dotted lines (minibands) are obtained by folding
of Si$_{50}$ (also H-Si$_{50}$) and Ge$_{50}$ (also H-Ge$_{50}$)
bands.}\label{fig:band_150}
\end{figure}
\end{center}

\begin{center}
\begin{figure}
\includegraphics[scale=0.53]{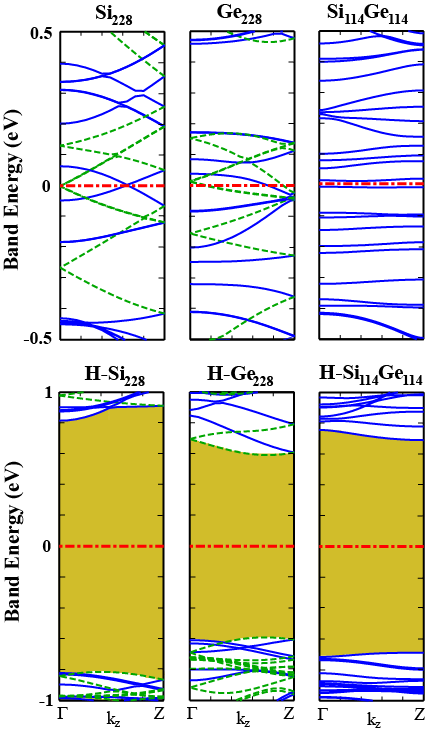}
\caption{(Color online) Energy band structures of optimized bare
and hydrogenated Si$_{2n}$, Ge$_{2n}$ nanowires and
Si$_{n}$Ge$_{n}$ nanowire superlattices for {\it n}=114. Zero of
energy is taken at the Fermi level. Dashed and dotted lines are
obtained by folding of Si$_{57}$ (also H-Si$_{57}$) and Ge$_{57}$
(also H-Ge$_{57}$) bands. }\label{fig:band_228}
\end{figure}
\end{center}

\begin{center}
\begin{figure}
\includegraphics[scale=0.43]{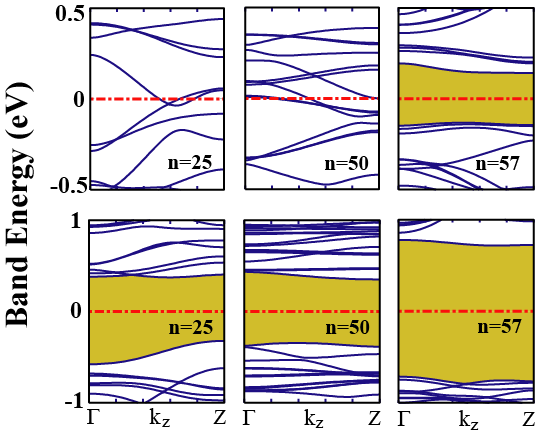}
\caption{(Color online) Energy band structure of bare and
hydrogenated Si$_{n}$Ge$_{n}$ nanowire superlattices for {\it
n}=25, 50 and 57. }\label{fig:triple_double_band}
\end{figure}
\end{center}

\section{Electronic Properties}
The band structures of optimized bare and hydrogenated
Si$_{n}$Ge$_{n}$ are given in Fig.\ref{fig:band_150} and
Fig.\ref{fig:band_228} for {\it n}=75 and 114, respectively. In
the same figures the band structures of bare and hydrogenated
Si$_{2n}$ and Ge$_{2n}$ constituent nanowires are presented for
the sake of comparison. Si$_{N}$ ({\it N}=25 and 57) and hence any
Si$_{2n}$ ({\it n=sN}) nanowires are metallic due to the surface
dangling bonds. Similarly Ge$_{N}$ ({\it N}=25 and 57) and hence
any Ge$_{2n}$ are metallic. Upon passivation of dangling bonds,
these metallic nanowires become semiconductor. For example,
H-Si$_{150}$ and H-Ge$_{150}$ nanowires have indirect band gaps,
$E_{g}$=1.1 eV and 0.7 eV, respectively. Normally, the band gap of
a H-Si$_{n}$ is inversely proportional to its diameter, if the
corresponding ideal nanowire cut from the bulk crystal were
directly passivated with H before the structural optimization.
Also the band gap is affected by the cross section geometry for
small {\it N}. For large {\it N}, the variation of $E_{g}$ with
{\it N} is more uniform.

Like Si$_{150}$ and Ge$_{150}$, Si$_{75}$Ge$_{75}$ is metallic.
The ideal equilibrium ballistic conductance of Si$_{150}$,
Ge$_{150}$ nanowires and Si$_{75}$Ge$_{75}$ NWSL is revealed to be
6$e^{2}/h$, 10$e^{2}/h$ and 8$e^{2}/h$, respectively. Since
H-Si$_{150}$ and H-Ge$_{150}$ are
semiconductors, H-Si$_{75}$Ge$_{75}$ NWSL is also semiconductor:
Its band gap is 0.7 eV and close to the band gap of H-Ge$_{150}$.
H-Si$_{114}$Ge$_{114}$ has a direct band gap of 1.4 eV. Again
it is smaller than the band gap of H-Si$_{228}$, but closer to
that of H-Ge$_{228}$.

In Fig.\ref{fig:triple_double_band} we examine how the electronic
energy bands of nanowire superlattices evolve with the lattice
constant $c$ or {\it s}. In the case of {\it N}=25, bare
Si$_{n}$Ge$_{n}$ nanowire superlattices are metallic for all {\it
n} ({\it n}=25, 50 or {\it s}=1, 2 and 3). As {\it s} increases
additional minibands occur and they become flatter. As for
H-Si$_{n}$Ge$_{n}$'s, they are all semiconductor for {\it n}=25,
50 and 75. As {\it n} increases all bands including lowest
conduction and highest valence band become flatter with the
formation of minibands. In this respect, the band gap becomes more
uniform as {\it s} increases. Similar behaviors are displayed also
for H-Si$_{n}$Ge$_{n}$ with {\it n}=57 and 114 (see
Fig.\ref{fig:band_228}). Unexpectedly, bare Si$_{n}$Ge$_{n}$'s are
semiconducting for {\it n}=57 and 114. The band gap decreases from
0.27 eV to 0.02 eV as {\it s} increases from 1 to 2. Isosurface
charge densities of these states near the band gap edges and found
that they are confined in one of the zones. It is concluded that
opening of the band gap originates from the mismatch of surface
dangling bond states in Si and Ge zones.

It should be noted that the band gap is underestimated by the GGA
calculations used in the present study. GW corrections performed
recently\cite{Zhao} for H-Si$_{n}$ in different orientation is in
the range of 0.5-0.6 eV for large diameters. In view of the fact
that Ge bulk is predicted as metal by GGA calculation, GW
correction for H-Ge$_{n}$ nanowires is expected to be in the same
range as that for H-Si$_{n}$. Under this circumstances a scissor
operation (namely increasing the band gap of corresponding
Si$_{n}$Ge$_{n}$ by the same amount 0.5-0.6 eV) may yield the
actual band gap. In summary, the band gaps of H-Si$_{n}$Ge$_{n}$
predicted by GGA calculation are underestimated, and actual bands
are expected to be 0.5-0.6 eV larger.

\begin{center}
\begin{figure}
\includegraphics[scale=0.44]{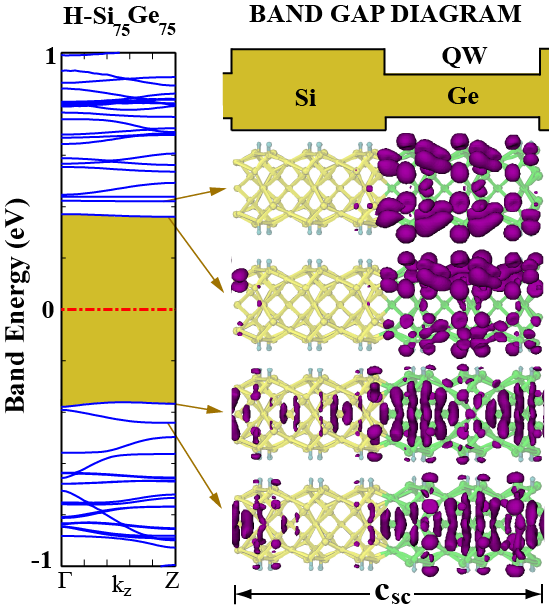}
\caption{(Color online) Schematic description of energy band
diagram in the unit cell (on the right and upper side), band
structure in momentum space (on the left side) and isosurface
charge density of states of H-Si$_{75}$Ge$_{75}$ NWSL at the band
edges. }\label{fig:band_charges_150}
\end{figure}
\end{center}

\begin{center}
\begin{figure}
\includegraphics[scale=0.44]{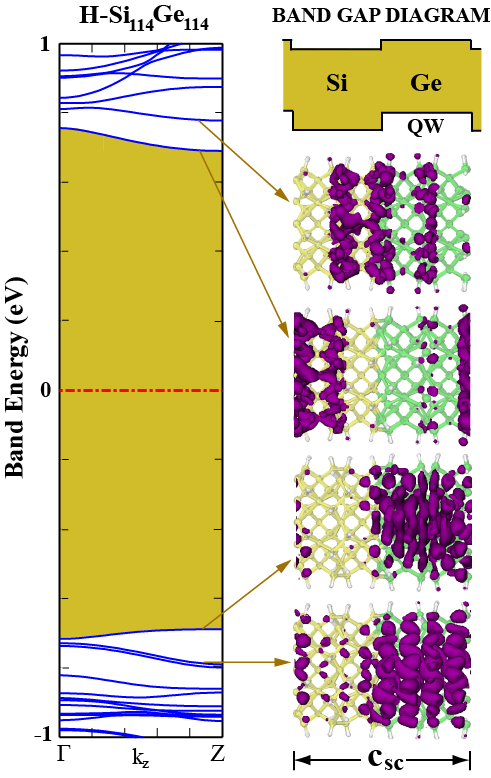}
\caption{(Color online) Schematic description of energy band
diagram in the unit cell (on the right and upper side),band
structure in the momentum space (on the left side) isosurface
charge density of states of H-Si$_{114}$Ge$_{114}$ NWSL at the
band edges. }\label{fig:band_charges_228}
\end{figure}
\end{center}

\section{confined states}
The results discussed in the previous section reveals that
Si$_{n}$ and Ge$_{n}$ nanowires making a Si/Ge heterojunction in
the supercell have band gaps of different width. Upon a
pseudomorphic junction the bands and hence band gaps corresponding
to Si and Ge zones are aligned. Combination of two features,
namely Si and Ge zones having different band gaps and band-lineup
result in band discontinuities and hence band-offsets. The
conduction and valence band edges of different zones (Si-zone or
Ge-zone) in the nanowire superlattice will have different
energies. Under these circumstances, the diagram of the conduction
band edge along the axis of NWSL will display a multiple quantum
well structure with the periodicity of $c_{sc}$ like a
Kronig-Penny model. Electrons in the well region of a zone should
decay in the adjacent zones having higher conduction band edge,
since their energy will fall into the band gap of this barrier
zone. As a result, the states of these confined (or localized)
electrons are propagating in the well, but decaying in the
barrier. Usually, confined electrons have low group velocity. They
may become more localized if the barrier is high and the width of
barrier is large. If the confinement (or localization) is
complete, the associated band E$_{n}$(k$_{z}$) becomes
flat.\cite{confined} Similar arguments are valid for the hole
states if the energies of valence band edges of both zones are
different.

In the past, the reference energies in determining band-offsets of
2D superlattices have been actively studied both experimentally
and theoretically. Energy diagram of conduction and valence band
edges are then used as effective potential forming a multiple
quantum well structure.\cite{Esaki} The states of conduction band
electrons and holes of valence band were treated using Effective
Mass Theory (EMT). These states are free electron-like 2D bands in
the planes and Bloch states forming minibands perpendicular to the
planes. The conditions are, however, different in NWSLs. First of
all, EMT may not be applicable directly in the present case, in
particular for NWSLs with small diameter. Secondly, the reference
energy level determined for planar superlattices may not be
appropriate. Recently, Kagimura {\it et al}\cite{Kagimura}
proposed surface dangling bond states as reference level for Si/Ge
core-shell superlattices. Under estimation of band gaps by DFT GGA
calculation may hinder the accurate determination of band-lineups.
Voon and Willatzen\cite{Voon} draw attention to the lateral
confinement of states in NWSL's. Using one-band EMT and by solving
Ben Daniel-Duke\cite{Duke} equation they found that the effective
barrier is lowered due to the coupling between radial and
longitudinal confinement. In particular, they predicted, that the
effective barrier and hence confinement disappears below a
critical radius of $\sim$ 5 nm. In the present study, the maximum
radius of NWSL was $\sim$ 0.9 \AA~ which is much lower than the
critical radius set for GaAs/AlGaAs NWSLs.\cite{Voon}

In the present study we examined whether some of states can be
longitudinally confined, by performing an extensive analysis of
charge densities of superlattice bands calculated by
first-principle methods. The formation of periodic quantum well
structure is schematically described in
Fig.\ref{fig:band_charges_150}. We expect that the values of band
gaps in the H-Si and H-Ge zones in a unit cell of the
H-Si$_{75}$Ge$_{75}$ cannot deviate significantly from the values
calculated for periodic H-Si$_{25}$ and H-Ge$_{25}$ nanowires
(namely, 1.1 eV and 0.7 eV, respectively). When the two zones are
connected by an atomically flat interface, H-Ge zone can form a
well between adjacent H-Si zones, since the band gap of the former
zone is smaller and the energy of conduction band edge is lower
relative to that of the latter zone. Upon normal band-lineup,
H-Ge$_{75}$ zone acts as a quantum well for both lowest conduction
and highest valence band electrons. Band structure of
H-Si$_{75}$Ge$_{75}$ with two lowest conduction and two highest
valence minibands and their isosurface charge distribution in the
superlattice unit cell are shown in
Fig.\ref{fig:band_charges_150}. The distribution of electronic
charge density is confirming the above normal band-lineup. Both
conduction band states are confined in the H-Ge$_{75}$ zone, but
they have very small weight in the H-Si$_{75}$ zone. Similarly,
states corresponding to two highest valence bands are also
confined in the H-Ge$_{75}$ zone. It should be noted that owing to
the charge transfer between adjacent zone the form of the energy
band diagram may change from the simple form given in
Fig.\ref{fig:band_charges_150}.

In Fig.\ref{fig:band_charges_228} we present similar analysis for H-Si$_{114}$Ge$_{114}$
NWSL. As compared to H-Si$_{75}$Ge$_{75}$, here H-Si and H-Ge
zones are $\sim$ 5 $\AA$ shorter. However, there are more
minibands owing to larger number of Si and Ge atoms. The ways the
highest valence band and the lowest conduction band states are
confined in different zones suggest a staggered band line-up.
Highest valence band states are confined in the H-Ge zone; but
lowest conduction band states are confined in H-Si zone.
States of 6$^{th}$ and 7$^{th}$ valence band
(from the top) are propagating throughout the NWSL.

We believe that present {\it ab-initio} results revealing confined
states for NWSLs with a radius as small as $\sim$ 0.6 nm are not
contradicting the conclusions obtained from one-band EMT model. We
think EMT as applied in Ref. 7 has to be revised for small
diameter NWSL.\cite{GaAs} We also note that H-Si$_{n}$Ge$_{n}$
nanowire superlattice has 1D rodlike structure. There are several
minibands in the 1D BZ. Number of minibands in a given energy
interval increases with either increasing {\it N} ({\it i.e.}
increasing diameter) or increasing {\it s}. A nanowire
superlattice with a long unit cell having several Si or Ge atoms
will have several (quasi continuous) minibands. States of
H-Si$_{n}$ or H-Ge$_{n}$ zone of the same energy are more likely
to match each other to construct a state that propagate throughout
the NWSL. Otherwise, a superlattice of small radius with short
unit cell have small number of bands. Then the states in different
zones are less likely to match. A state, which cannot find a
matching partner is confined to its zone. As a matter of fact, we
were able to deduce confined states even in the barrier zone
(H-Si) with energies higher than the conduction band edge.

\section{Conclusion}
Atomic structure of H-Si$_{n}$ and H-Ge$_{n}$ nanowires is
tetrahedrally coordinated near the center, but at the surface
deviates significantly from corresponding bulk crystal. Calculated
force constants indicate that Si$_{n}$ is stiffer than Ge$_{n}$.
Generally nanowires become stiffer after passivation with
hydrogen. These two nanowires are 1D semiconductors with their
band gap depending on their diameter and also on the geometry of
their relaxed cross section. If finite segments of these nanowires
are joined pseudomorphically and the resulting heterostructure are
repeated periodically along the axis of the wires, one obtains a
H-Si$_{n}$Ge$_{n}$ superlattice structure. In these longitudinal
NWSLs electrons are normally bound to the wire in radial
direction, but propagate along their axis. A specific state which
propagates in one zone (say H-Si) can decay in the adjacent zone
(say H-Ge), when a matching state in the same energy is absent.
Such a state is called confined state. Our charge density analysis
indicate that Si/Ge NWSL with radius as small as 0.6 nm can have
confined states at the band edges and also within the conduction
and valence band. Confined states offer interesting device
applications. NWSL has an important advantage that the device part
and leads can be produced from similar nanowires. Theoretically,
NWSL have several interesting issues to be clarified. In
particular, theories derived from planar superlattices to predict
band-lineups and model calculations using EMT have to be revised
for small diameter NWSLs.


\begin{thebibliography}{99}
\bibitem{strain} These calculations have been performed by minimizing the total energy under a preset uniaxial strain in the
elastic range. The strain on Ge and Si-zone calculated are
determined after full relaxation.
\bibitem{GaAs}  We also note that the material parameters of GaAs/AlGaAs relevant for EMT by themselves are different
from Si$_{n}$Ge$_{n}$ nanowire superlattices studied here.
\end{thebibliography}
\end{document}